\documentclass[aps,twocolumn,nofootinbib]{revtex4}
\usepackage{mathrsfs} \usepackage{amsmath} \usepackage{graphicx}
\usepackage{dcolumn} \usepackage{bm} \usepackage{amssymb}
\usepackage{latexsym}
\def\theequation{\arabic{section}.\arabic{equation}}
\newcommand{\be}{\begin{equation}}
\newcommand{\ee}{\end{equation}}
\begin{document}
\title{An analysis of the Sultana-Dyer cosmological black hole 
solution of the Einstein equations}
\author{Valerio Faraoni} \email{vfaraoni@ubishops.ca}
\affiliation{Physics Department, Bishop's University, 2600
College Street, Sherbrooke, Qu\'{e}bec, Canada J1M~1Z7}

\date{}

\begin{abstract}
The Sultana-Dyer solution of general relativity representing a 
black hole embedded in a special cosmological background is 
analysed. We find an expanding (weak) spacetime singularity 
instead of the reported  conformal Killing horizon, which is 
covered by an expanding black hole apparent horizon (internal to 
a  cosmological apparent horizon) for most of the history of the 
universe. This singularity was naked early on. The global 
structure of the solution is studied as well.
\end{abstract}

\maketitle

\section{Introduction}
\setcounter{equation}{0}

Two main areas of gravitational physics, cosmology and the 
study of black holes, come together in the attempt to find exact 
solutions of the Einstein equations describing black holes 
which are asymptotically Friedmann-Lemaitre-Robertson-Walker 
(FLRW) instead of asymptotically flat.  With the exception of 
static solutions, such as the Schwarzschild-de Sitter black hole 
or its generalizations, these exact solutions describe dynamical 
black holes. What is meant exactly by ``black hole'' 
when the metric is non-stationary and the  usual 
teleological event horizon is not present is a non-trivial 
question which  has led to extensive research on dynamical 
horizons and their mechanics and thermodynamics ({\em e.g.}, 
\cite{Nielsen, AshtekarKrishnan, Booth, Visser} and references 
therein). The well-known black hole thermodynamics \cite{Wald} 
hinges on Hawking's discovery of thermal radiation from black 
holes \cite{Hawkingradiation}, and this calculation relies on 
neglecting the backreaction of radiation on the spacetime 
metric. The full treatment of backreaction and a 
full understanding of time-dependent dynamical horizons are still 
in the future. It interesting, therefore, to find and study  
exact solutions of the Einstein field equations describing black 
holes, 
by which  we mean central singularities covered by an {\em 
apparent} horizon-these can be used as testbeds for various 
theoretical characterizations  of surface gravity, energy 
fluxes, {\em etc}. Only a 
few such solutions are known. A 
complication arising when a black hole is somehow embedded in a 
cosmological background other than the de (anti-)Sitter one 
sourced by  a cosmological constant, is that the cosmic fluid 
tends to accrete onto the central black hole. Forbidding this 
accretion flow results in a rather artificial condition and in 
the old McVittie solution \cite{McVittie}. 

There are other motivations for pursuing exact cosmological black 
hole spacetimes: one is the problem of the effect of the 
cosmological expansion on local dynamics (\cite{CarreraGiulini} 
and references therein), which generated  
the McVittie solution \cite{McVittie}. It has been realized 
\cite{Price} that 
participation in the expansion of the cosmic 
substratum may be more difficult to achieve for strongly  
than for 
weakly bound local systems, hence one wants to look at the 
most strongly bound local system, the black hole. This approach 
has led to the solutions of Sultana and Dyer \cite{SultanaDyer} 
and those of \cite{McClureDyer, FaraoniJacques, fate}. Recent 
interest in phantom dark energy and its properties (including 
thermodynamics)  \cite{Babichevetal, SaidaHaradaMaeda} 
motivated the study of the  backreaction due 
to the accretion of cosmological phantom energy  onto black 
holes and the possibility that phantom energy may violate  
Cosmic Censorship \cite{Gaoetal}. Exact solutions describing 
black holes in a cosmological fluid may also lead to toy models 
for  
evaporating black holes.  Finally, alternative 
gravitational theories 
such as $f(R)$ gravity have received much attention recently as 
possible models of the present acceleration of the universe 
without postulating exotic dark energy (\cite{CCT}, see 
\cite{review} for a 
review and \cite{otherreviews} for shorter introductions to this  
subject). $f(R)$ gravity and all 
theories of modified gravity introduced for this purpose produce 
an effective time-varying cosmological constant and 
spherically symmetric solutions in these theories are not likely 
to be asymptotically flat but rather asymptotically FLRW. Black 
holes 
will not be Schwarzschild-like but dynamical: in fact, the 
Jebsen-Birkhoff theorem is not valid in these theories and 
spherically symmetric solutions will, in general, be 
non-stationary (see, for example, the dynamical solution proposed 
in 
\cite{Clifton} in the context of metric $f(R)=R^{1+\delta}$ 
gravity). The interest in cosmological black holes is not 
confined to this class of alternative theories (which are, 
anyway, special cases of scalar-tensor gravity \cite{review, 
otherreviews}): interest has come from the 
possibility that inhomogeneities lead to local variations 
of the effective gravitational constant in scalar-tensor 
cosmologies \cite{CliftonMotaBarrow}. Exact cosmological and 
time-dependent  black
holes are of interest also in higher-dimensional Gauss-Bonnet 
gravity \cite{HidekiGaussBonnet} and arise from 
intersecting 
branes in supergravity \cite{KMaeda}.

In spite of all these motivations, only a handful of exact 
solutions describing dynamical black holes embedded in   
cosmological backgrounds are known, and even fewer are 
properly understood. Here 
we analyse the Sultana-Dyer solution \cite{SultanaDyer} which 
reserves a few suprises, and discuss its singularities, apparent 
horizons, global structure, and physical interpretation.  We 
adopt the notations of Ref.~\cite{Wald}.

\section{The Sultana-Dyer solution}
\setcounter{equation}{0}

The Sultana and Dyer solution \cite{SultanaDyer} is a metric  
of  Petrov type~D interpreted by its discoverers as a black hole 
embedded in a spatially  flat FLRW universe with scale factor $ 
a(t)\propto t^{2/3}$ (where $t$ is the  comoving time). It is 
generated by  conformally transforming the Schwarzschild metric 
$g_{ab}^{(S)} \rightarrow g_{ab}= \Omega^2 \, g_{ab}^{(S)}$  with 
the  intention of  of mapping  the Schwarzschild timelike Killing 
field  $\xi^c$ into a  conformal Killing field (for 
$\xi^c \nabla_c \Omega \neq 0$), and the Schwarzschild 
event horizon into a conformal Killing 
horizon. The choice   $\Omega=a(t)=a_0 \, t^{2/3}$  (the scale 
factor of a  dust-filled $k=0$ FLRW universe) generates the 
metric in the  form given by Sultana and Dyer \cite{SultanaDyer}
\begin{eqnarray}
ds^2 &= &  a^2(\eta) \left[ -\left( 1-\frac{2m}{\tilde{r}} 
\right) d\eta^2 
+ \frac{4m}{\tilde{r}}\, d\eta d\tilde{r} 
\right.\nonumber\\
&&\nonumber\\
&+& \left. \left( 
1 +  \frac{2m}{\tilde{r}} \right) \, d\tilde{r}^2 
+\tilde{r}^2d\Omega^2 \right] \;,\label{SultanaDyeroriginal}
\end{eqnarray}
where $m$ is a constant, $d\Omega^2=d\theta^2+\sin^2 \theta 
d\varphi^2 $ 
is the metric on the unit 2-sphere, $a(\eta)=a_0 \eta^2= a_0 
t^{2/3}$. We use the metric signature opposite to that of 
\cite{SultanaDyer} and the symbol $\eta$ and $t$ for conformal 
and  
comoving time, respectively, with $dt=ad\eta$  (these notations 
are switched with 
respect to those of \cite{SultanaDyer} but follow standard 
practice in cosmology). 

The Sultana-Dyer metric~(\ref{SultanaDyeroriginal}) is obtained 
by 
conformally transforming the Schwarzschild metric written in 
Painlev\'e-Gullstrand coordinates. Of course, the line 
element~(\ref{SultanaDyeroriginal}) can be cast in a  form 
explicitly conformal to Schwarzschild written in the more 
familiar Schwarzschild coordinates.  By introducing the new time 
coordinate $\bar{t}$ defined by
\be
dt=d\bar{t}+\frac{2ma\, d\tilde{r}}{\tilde{r}\left( 
1-\frac{2m}{\tilde{r}} \right)} \;,
\ee
the line element~(\ref{SultanaDyeroriginal}) transforms to 
\begin{eqnarray}
ds^2& =& -\left( 1-\frac{2m}{\tilde{r}} \right) d\bar{t}^2 
+\frac{a^2 \,d\tilde{r}^2 }{1-\frac{2m }{\tilde{r}} }   +a^2 
\tilde{r}^2 d\Omega^2 \label{2} \nonumber\\
&&\nonumber\\
&= & a^2\left[ 
-\left( 1-\frac{2m}{\tilde{r}} \right) d\bar{\eta}^2 
+\frac{d\tilde{r}^2}{1-\frac{2m}{\tilde{r}} } + \tilde{r}^2 
d\Omega^2 \right] \nonumber\\
&& \label{3}
\end{eqnarray}
which is manifestly conformal to Schwarzschild with conformal 
factor $a$ and $ d\bar{t} = a d\bar{\eta}$. It is obvious 
that it reduces to a spatially flat 
FLRW  metric as $r\rightarrow +\infty$.

By using the   isotropic radius $r$ defined by 
\be
\tilde{r}= r\left( 1+\frac{m}{2r} \right)^2 \;,
\ee
the  line element~(\ref{3}) becomes
\be 
ds^2 =  - \frac{  \left(1-\frac{m}{2 r} \right)^2}{
\left(1+\frac{m}{2 r} \right)^2} \, d\bar{t}^2+
a^2 \left( 1+\frac{m}{2r} \right)^4 \left( 
dr^2 +r^2 d \Omega^2 \right)  \label{SDisotropic}
\ee
in isotropic coordinates $\left( \bar{t}, r, \theta, \varphi 
\right)$. 
The metric~(\ref{SDisotropic})  is 
recognized to be formally the same as the McVittie 
solution \cite{McVittie} but 
with the important difference that the  metric coefficient $m$ is 
now a constant, contrary to the McVittie case in which 
$  \, d m/md\bar{t} =- \,  
da/ ad\bar{t}  $. We will 
refer to this equation  as the ``McVittie no accretion 
condition''. The latter 
stems from the explicit requirement  that there is no accretion 
of cosmic fluid onto the central object. In modern language, this 
has been recognized as the constancy of the Hawking-Hayward 
quasi-local 
mass $M \equiv m( \bar{t})a( \bar{t})$ \cite{NolanCQG}. Since the 
McVittie no 
accretion condition is violated by the Sultana-Dyer solution, the 
latter describes an accreting object, as Sultana and Dyer 
recognize from inspection of the field equations and 
the fact that $T_0^1 \neq 0$. 

A second important difference between the Sultana-Dyer and the 
McVittie metric is that the latter is sourced by a  single 
perfect fluid while the material source of the former is a 
mixture  of two non-interacting perfect fluids, a null dust and a 
massive  dust \cite{SultanaDyer}.  The stress-energy tensor is 
\be
T_{ab}= T_{ab}^{(I)}+T_{ab}^{(II)} \;,
\ee
where $ T_{ab}^{(I)}=\rho \, u_a\, u_b $ describes an ordinary 
massive dust and $ T_{ab}^{(II)}=\rho_n \, k_a\, k_b$ 
describes a null dust with density $\rho_n $ and $k^c k_c=0$ 
\cite{SultanaDyer}. The use of two fluids instead of one 
follows
from the fact that, once accretion is allowed and the energy 
density and pressure depend on the radial, as well as the time,  
coordinate, solutions sourced by a single perfect fluid do not 
exist, 
except for the Schwarzschild-de Sitter metric (which is a special 
case of the McVittie solution) \cite{Gaoetal,FaraoniJacques}.

We will make use of the quantity 
\be\label{bigM}
 M(\bar{t}) \equiv m \, a( \bar{t}) \;,
\ee
which is not constant in the Sultana-Dyer solution. These authors  
identify a conformal Killing horizon, the locus 
$\tilde{r}=2m$ (or $r=m/2$ in terms of the isotropic radius 
$r$) 
\cite{SultanaDyer}. 
This is obtained by mapping the Schwarzschild 
event horizon of the seed metric (a Killing horizon) into the 
2-surface $\tilde{r}=2m$. Therefore, Sultana and Dyer 
interpret their metric as describing a  
black hole embedded in a  special (spatially flat) FLRW 
background 
universe. The authors identify certain problems with this 
solution: the accretion flow onto the black hole becomes 
superluminal, and the energy density of the cosmic fluid becomes 
negative after a certain time near the conformal Killing horizon 
at $\tilde{r}=2m$. They proceed to study radial null geodesics 
and surface gravity on this surface, and the behaviour of 
timelike geodesics representing, {\em e.g.}, planetary orbits, in 
this metric. This is inspired by the old problem of general 
relativity of whether the cosmic expansion affects local systems 
(see~\cite{CarreraGiulini} for a recent review).

Here we take the study of the Sultana-Dyer solution a 
step further. We find that it is quite different from the picture 
provided by these authors: in particular, we show that the locus 
$\tilde{r}=2m$ is not a conformal Killing horizon but a spacetime 
singularity. It is a weak singularity in the sense of Tipler's 
classification \cite{Tipler}. If the current interpretation of 
the McVittie metric \cite{NolanCQG} applies to the Sultana-Dyer 
solution, the two disconnected regions $\tilde{r}>2m$ and $0< 
\tilde{r}<2m$  correspond to a black hole and white hole region, 
respectively. The spacetime singularity $\tilde{r}=2m$ is covered 
by an apparent horizon, which we locate together with a 
cosmological horizon. This alleviates the 
problems of negative energy density and superluminal flux near 
$\tilde{r}=2m$. The singularity expands with time, comoving with 
the cosmic substratum but the horizon expands at a  slightly 
smaller rate and, in the infinite future, it approaches 
the singularity asymptotically. The $\tilde{r}>2m$ 
Sultana-Dyer spacetime has also a Big Bang spacelike singularity.  
In the early universe, near the Big Bang, the $\tilde{r}=2m$ 
singularity was naked, and got covered by an apparent horizon 
only later. All these features resemble certain  
characteristics  of 
the Fonarev solution \cite{Fonarev, HidekiFonarev} or of  
generalized McVittie solutions \cite{FaraoniJacques, Gaoetal, 
fate}.

\section{Singularities and apparent horizons of the Sultana-Dyer 
solution}
\setcounter{equation}{0}

Let us begin by examining the locus $\tilde{r}=2m$: this is not a  
conformal Killing  horizon as stated in \cite{SultanaDyer}, but a 
spacetime singularity.  In fact, the Ricci curvature is 
\be
{R^a}_a=\frac{12}{a^2(\bar{\eta})\left( 
1-\frac{2m}{\tilde{r}}\right)} 
=\frac{6\left( \partial_{\bar{t}} \bar{H} +2 \bar{H}^2 
\right)}{1-\frac{2m}{\tilde{r}} } 
\label{ssscurvature} 
\ee 
(where $\bar{H}=d  \ln a/ d\bar{t}$)  and it diverges as 
$r\rightarrow m/2$ (the discrepancy with the 
Sultana-Dyer paper is discussed in Appendix~A). This is a 
covariant statement and not an artifact of the coordinate 
system adopted because the Ricci scalar is an 
invariant of the Riemann tensor. ${R^a}_a$ also diverges as 
$a\rightarrow 0$, the usual Big Bang singularity of the  
background FLRW universe. The metric determinant  is
\be
g=-a^8( \bar{\eta}) \, \tilde{r}^4  \sin^2 \theta
\ee
and does not tend to zero as $\tilde{r}\rightarrow 2m$, hence 
an object does not get crushed to zero volume as it 
approaches this singularity, which is a weak one  in  Tipler's 
classification \cite{Tipler}. It will be shown 
below that this singularity is covered by an apparent horizon 
and therefore, as in the McVittie 
metric, the regions $\tilde{r}>2m$ and $0< \tilde{r}<2m$ may be 
interpreted as describing a black hole and  a white hole region 
\cite{NolanCQG}. Although this interpretation does  not seem 
absolutely compelling to us in view of the fact that a 
physical object or particle  could potentially cross a weak 
singularity, we do not address this issue here and refer the 
reader to the comprehensive discussion of Nolan \cite{NolanCQG} 
and to the references therein.

The issue is now to determine whether the $\tilde{r}=2m$ 
singularity is naked or covered by an apparent horizon and, 
therefore, whether it really describes a black hole embedded in a  
cosmological background. Since the metric is non-stationary, 
there is no event horizon and the appropriate horizon notion is 
that of apparent horizon \cite{Nielsen}.  We proceed to rewrite 
the Sultana-Dyer 
solution in the Nolan gauge, in which it will be straightforward 
to decide whether apparent horizons exist and, if so, to locate 
them. 

Using the area radius  $ R\equiv a\tilde{r}$, which is a 
geometric quantity, eq.~(\ref{bigM}), and the 
fact that
\be
d\tilde{r}=\frac{dR}{a}-HRd\eta \;,
\ee
where $H\equiv \dot{a}/a$ and an overdot denotes 
differentiation with respect to $t$, the  
line element~(\ref{SultanaDyeroriginal}) becomes 
\begin{eqnarray}
ds^2 &= & -\left( 1-\frac{2M}{R}-\frac{  H^2R^2}{ 1-\frac{2M}{R}} 
\right)dt^2 +\frac{dR^2}{1-\frac{2M}{R}} \nonumber\\
&&\nonumber\\ 
&-& \frac{2HR \, dtdR}{1-\frac{2M}{R}}  + R^2 d\Omega^2 
\label{intermediate} 
\end{eqnarray}
in coordinates $\left( t, R, \theta, \varphi \right)$, with the 
singularity located at $R=2M$. 
Let us use $A \equiv 1-2M/R$ and the new 
time  coordinate $T$ defined by
\be
dT= \frac{1}{F} \left( dt +\frac{HR}{ A^2-H^2R^2 }\, dR \right)
\ee
where $F(T(t,R), R)$ is an integrating factor satisfying
\be
\frac{\partial}{\partial R}\left( \frac{1}{F} \right)= 
\frac{\partial}{\partial t} \left( \frac{HR}{F\left( A^2-H^2R^2 
\right)} \right)
\ee
to guarantee that $dT$ is an exact differential. After 
straightforward manipulations the line 
element~(\ref{intermediate}) is   recast in the Nolan 
gauge as
\begin{eqnarray}
ds^2 &= & -\left( 1-\frac{2M}{R}- \frac{H^2R^2}{1-\frac{2M}{R}} 
\right) F^2 dT^2 \nonumber\\
&&\nonumber\\   
&+ & \frac{dR^2}{1-\frac{2M}{R} -\frac{H^2R^2}{1-\frac{2M}{R} } } 
  +  R^2 d\Omega^2 \;.\label{SultanaDyerNolangauge}
\end{eqnarray}
The apparent horizons, if they exist, are the locus $g^{RR}=0$, 
or 
\be \label{locateAHs}
HR=\pm\left( 1-\frac{2M}{R}  \right)\;.
\ee
We discard the lower  sign in eq.~(\ref{locateAHs}) which 
corresponds to a  contracting universe, and the apparent 
black hole and cosmological horizons are given by 
\be
R_{bh}(t) =\frac{1-\sqrt{ 1-8MH}}{2H} \;, \;\;\;\;\;
R_{c}(t)=\frac{1+\sqrt{ 1-8MH}}{2H} \;,
\ee
respectively. 
It must be $M \leq \frac{H^{-1}}{8}$, or $ \dot{a} \leq 
\frac{1}{8m}$ for these apparent horizons to 
exist. In the limit $m\rightarrow 0$ the black hole 
and its apparent horizon disappear and the cosmological apparent 
horizon  
reduces to the familiar surface $R=\frac{1}{H} $. The fact that 
the 
proper radius of the latter is larger than $R_c$ can be 
interpreted as the pull of the central black hole on the cosmic 
fluid. 

Since  $  R(1-HR)=2M>0$ 
at the apparent horizons and $H>0$    in an expanding universe, 
it is 
\be
R_{bh}= \frac{2M}{1-HR_{bh}}> 2M 
\ee
and  the singularity $R=2M(t)=2ma(t)$, at which the 
 flow is superluminal and the energy density becomes 
negative-definite, is  hidden by the apparent 
horizon. This alleviates somehow the problems of this solution 
reported in \cite{SultanaDyer}.  However, since $a(t)=a_0t^{2/3}$ 
in  comoving time,  as $t 
\rightarrow +\infty$, $MH=m \dot{a} =\frac{2ma_0 }{3t^{1/3}}  
\rightarrow 0$ and  $R_{bh}\rightarrow 2M$  
approaching the singularity asymptotically.

By using the scale factor $a(t)=a_0t^{2/3}$ of the Sultana-Dyer 
metric it follows that the 
constant $m$ must satisfy $ m \leq \frac{3t^{1/3}}{16 a_0}$ for 
the apparent horizons to exist. This 
condition is violated at early times, implying that the $R=2M$ 
singularity  was naked at early times 
and later on (at $t   = \left( \frac{16a_0}{m} \right)^3$)  an 
apparent 
horizon appeared that immediately bifurcated into a 
cosmological horizon and a black hole apparent horizon  
covering the singularity.

Tha gobal structure of the Sultana-Dyer solution is analyzed in 
the next section.  To conclude this section, we discuss the 
Misner-Sharp \cite{MisnerSharp} and Hawking-Hayward 
\cite{HawkingHayward, Hayward} quasi-local energies.

The Misner-Sharp mass $M_{MS}$ is defined in terms of the area 
radius $R$ by $ 1-\frac{2M_{MS}}{R}=-\nabla^c R\nabla_c R$ 
\cite{MisnerSharp} which, at the black hole apparent 
horizon, yields
\be
M_{MS}=\frac{R_{bh}}{2}=\frac{1-\sqrt{1-8MH}}{4H} \;.
\ee
In order to compute the Hawking-Hayward quasi-local mass we 
introduce the null coordinates $\left( u, v \right)$ defined by
\begin{eqnarray}
du &= & \frac{1}{\sqrt{2}} 
\left[ \sqrt{ 
1-\frac{2M}{R} -\frac{H^2R^2}{ 1-\frac{2M}{R} } 
} \, F \, dT  \right.\nonumber\\
&&\nonumber\\
&-& \left. \frac{dR}{ 
\sqrt{  1-\frac{2M}{R} -\frac{H^2R^2}{1-\frac{2M}{R}} 
}
} \right] \;,\\
&&\nonumber\\
dv &=& \frac{1}{\sqrt{2}} 
\left[ \sqrt{ 
1-\frac{2M}{R}-\frac{H^2R^2}{1-\frac{2M}{R}}  
} \, F \, dT \right. \nonumber\\
&&\nonumber\\
& + & \left. \frac{dR}{
\sqrt{
1-\frac{2M}{R} -\frac{H^2R^2}{1-\frac{2M}{R}} 
}
} 
\right] \;,
\end{eqnarray}
in terms of which the metric is reduced to Hayward's standard 
form \cite{Hayward} 
\be
ds^2=-2 \,du \,dv +R^2 d\Omega^2 \;.
\ee
Since $dR=\sqrt{ 1-\frac{2M}{R}-\frac{H^2R^2}{1-\frac{2M}{R}}} 
\frac{\left(dv-du \right)}{\sqrt{2}}$, the Hawking-Hayward 
quasi-local mass  is easily computed using the 
prescription for spherical symmetry \cite{Hayward}
\be
M_{HH}=R\left( R_u R_v +\frac{1}{2} \right)
\ee
which yields, at the black hole apparent horizon,
\be
M_{HH}=M_{MS}=\frac{R_{bh}}{2}
\ee
and coincides with the Misner-Sharp mass. Both masses diverge in 
the limit $R\rightarrow 2M$.

\section{Global structure}
\setcounter{equation}{0}

The causal nature of the singularity and the 
black hole apparent horizon are determined as follows. 
The singularity is characterized by the equation $R-2M=0$ and the 
normal is obtained by taking the gradient of this equation (the 
limit 
$R\rightarrow 2M$ is implicit in the following). Unfortunately, 
the integrating factor $F$ appearing in the line  
element~(\ref{SultanaDyerNolangauge}) is not determined 
explicitly, hence it is more convenienent to use the coordinates  
$\left( t,R, \theta, \varphi\right)$ in which the metric 
 is given by eq.~(\ref{intermediate}) and its inverse 
by
\be
\left( g^{ab} \right)=\left(
\begin{array}{cccc}
\frac{-1}{1-\frac{2M}{R}} & \frac{-HR}{1-\frac{2M}{R}} & 0& 0 \\
& & &\\
\frac{-HR}{1-\frac{2M}{R}} & 
\left(1-\frac{2M}{R}-\frac{H^2R^2}{1-\frac{2M}{R}}\right) & 0 & 
\\
 & & & \\
0 & 0 & \frac{1}{R^2} & 0 \\
& & & \\
0 & 0 & 0 & \frac{1}{R^2\sin^2\theta } 
\end{array} \right) 
\ee
One obtains $ n_a=\nabla_a R -2m\dot{a} \delta_{a0}=\left( -2MH, 
1, 0,0  \right) $ and 
\be
n^a=g^{ab}n_b=
\left(  -HR , -H^2 R^2 +  
1- \frac{2M}{R} , 0, 0 \right) 
\ee 
and, in the limit $R -2M \rightarrow 0$ $n_a=\left( -2MH, 1, 0, 0 
\right)$, $n^a= \left( -2MH, -4M^2 H^2 , 0,0 \right)$, while 
$n^cn_c=0$. The $R=2M$ singularity is null, which is not 
surprising if one remembers that is is obtained by conformal 
mapping of the null event horizon of the Schwarzschild black 
hole  used as the seed metric in this solution-generating 
technique \cite{SultanaDyer}.

The black hole apparent horizon has equation $ f(t, R)=2RH - 1 
+ \sqrt{1-8MH} =0 $. The normal has the direction of the 
gradient $\nabla_a f$, or
\be
n_a=  \left(  B(t,R) , 2H, 0 ,0 \right)
\ee
where 
\be
B(t,R)=2R\dot{H} -\frac{4M \left( H^2 +\dot{H} \right)}{
\sqrt{1-8MH}} \;,
\ee
while  
\begin{eqnarray}
&& n^a =  g^{ab}n_b = \left( 
- \frac{\left( B+2H^2R \right)}{1-\frac{2M}{R}} ,   
\right. \nonumber\\
&&\nonumber\\
& &  \left.  \frac{-HRB}{1-\frac{2M}{R}}   +2H\left( 
1-\frac{2M}{R}-\frac{H^2R^2}{1-\frac{2M}{R}} \right),  0, 0 
\right) \;.
\end{eqnarray}
Therefore, 
\begin{eqnarray}
&& n^a n_a= -\frac{1}{1-\frac{2M}{R_{bh}}} \left\{
\left[ 2\dot{H}R_{bh}-\frac{ 4M\left( H^2+\dot{H} 
\right)}{\sqrt{1-8MH}} \right]^2 \right. \nonumber\\
&&\nonumber\\
&& \left. +   4H^2R 
\left[ 2\dot{H}R_{bh}-\frac{ 4M\left( H^2+\dot{H} 
\right)}{\sqrt{1-8MH}} \right] \right\} \nonumber\\
&&\nonumber\\
&& =
-\frac{1}{1-\frac{2M}{R_{bh}}} \left(C^2+4H^2R \right) 
\nonumber\\
&&\nonumber\\
&&= 
-\frac{1}{1-\frac{2M}{R_{bh}}} \left[ \left(C 
+2H^2R\right)^2 -4H^4R^2 \right] \nonumber\\
&&\nonumber\\
&&= \frac{\left|C \right|}{HR_{bh}}  \left(C 
+4H^2 R\right) \;,
\end{eqnarray}
where
\be
C= \frac{  \frac{-\dot{H}}{2}\left( 1-\sqrt{1-8MH} \right)^2 
-4MH^3}{H\sqrt{1-8MH}} \;. 
\ee
Now, 
\begin{eqnarray}
&& C +4H^2 R = \frac{1}{H\sqrt{1-8MH}} \left[
\frac{-\dot{H}}{2}\left( 1-\sqrt{1-8MH} \right)^2  \right.
\nonumber\\
&&\nonumber\\
&&\left. +12MH^3 -2H^2 
\left( 1-\sqrt{1-8MH} \right) \right] \label{questa} 
\end{eqnarray}
and at late times $H=\frac{2}{3t}$ and $\dot{H}=\frac{-2}{3t^2}$,  
hence the numerator of eq.~(\ref{questa}) is dominated by the 
term 
$12MH^3$  and $n^a n_a>0$: the black hole apparent horizon is 
timelike  at late times. At early times, but after the 
horizons have appeared, it is
\begin{eqnarray}
&& \frac{1}{ H\sqrt{1-8MH}} \left[
\frac{-\dot{H}}{2}\left( 1-\sqrt{1-8MH} \right)^2 +12MH^3 
\right.\nonumber\\
&&\nonumber\\
&& \left. -2H^2 
\left( 1-\sqrt{1-8MH} \right)\right] \nonumber\\
&&\nonumber\\
&& \simeq 
\frac{12ma_0}{t^{7/3} } \, \frac{1}{H\sqrt{1-8MH}} >0
\end{eqnarray}
and the horizon is timelike early on.

The global picture of the Sultana-Dyer spacetime that emerges 
from these considerations  is that of a universe with a  
spacelike  Big Bang singularity at $t=0$,  the null 
singularity $R=2M$ (the conformal mapping of a Schwarzschild 
event horizon) which was naked at early times, until it got 
covered  by a timelike apparent horizon appearing at a finite 
time. The conformal diagram is sketched in Fig.~1.

\begin{figure} 
\includegraphics[width=6.5cm]{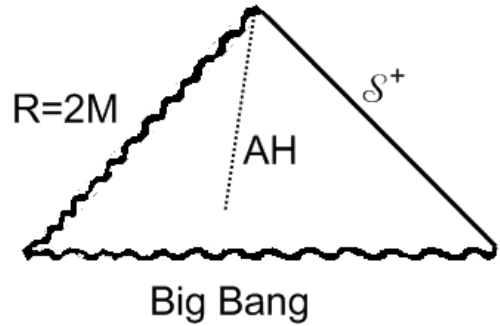}
\caption{The conformal diagram of the Sultana-Dyer spacetime. 
The horizontal wiggly line at the bottom describes the Big Bang 
singularity, the wiggly line at 45 degrees denotes the 
$R=2M$  null singularity, and the solid line at 45 degrees 
describes future null infinity. Null geodesics end at future null 
infinity 
or at the black singularity (either when it is naked if started 
early on, or crossing the timelike black hole apparent horizon 
labelled AH).}  
\end{figure}

\section{Discussion and conclusions}
\setcounter{equation}{0}

There are many motivations, discussed in the Introduction, to 
find and study exact solutions of the field equations of general 
relativity or alternative gravity theories 
describing dynamical black holes embedded in a cosmological 
background. Among these the Sultana-Dyer 
metric deserves some attention in view of its relative 
simplicity and of the technique used to generate it, which may 
lead to a  wider class of solutions. There is evidence that 
among the class of  generalized McVittie solutions 
\cite{FaraoniJacques, Gaoetal} those conformal to the 
Schwarzschild spacetime  (``comoving solutions'') are  generic 
\cite{fate}, and this constitutes further motivation to 
understand solutions seeded by the Schwarzschild 
metric, such as the Sultana-Dyer one.

To summarize and discuss the salient features of the  
Sultana-Dyer spacetime, we found that the 
conformal image of the Schwarzschild event horizon  is not 
 a conformal Killing horizon but rather a spacetime 
singularity at $R=2m$ (where $R$ is the area radius). This 
singularity is weak in Tipler's sense and is null, as 
should be expected because  the  seed Schwarzschild event horizon 
is a 
null surface. This expanding singularity  is 
interpreted as the effect of the pull of the cosmological matter 
stretching the $r=0$ singularity of the Schwarzschild spacetime 
into 
one of finit radial extent. Early  on, this singularity is naked 
and only later a timelike  apparent horizon develops 
which immediately bifurcates into a black hole apparent horizon 
covering the singularity and an apparent cosmological horizon. 
The radius of the cosmological horizon is smaller than the value 
$ H^{-1}$ of the Hubble radius in a  FLRW spacetime, which may be 
interpreted as an effect due to the gravitational pull of the 
central object on the cosmic fluid.

The singularity expands comoving with the cosmic substratum, 
while the apparent horizon expands at a slightly smaller rate 
and eventually comes to coincide with the singularity in the 
infinite future. Sultana and Dyer report superluminal flow near 
the singularity, and one may question whether the notion of black 
hole makes sense at all in the presence of superluminal flows 
which, in principle, allow particles to escape from the 
apparent horizon. However, it should be noted that this 
radial flow is always directed inward and nothing actually 
escapes from the black hole apparent  horizon. In our opinion, 
this unpleasant superluminal 
feature is due to the simplicity of the model under study and 
will not be present in  more realistic and sophisticated  models 
yet to come. 
For the moment we content ourselves with understanding the simple 
model~(\ref{SultanaDyeroriginal}). The Sultana-Dyer spacetime 
exhibits  the spacelike Big 
Bang cosmological singularity, the null black hole singularity  
and, later on, the two apparent horizons. As is clear from the 
conformal diagram 
of Fig.~1, the timelike black hole apparent horizon meets the 
null singularity in the infinite future, with the dust-dominated  
universe expanding and diluting forever.

In addition to the superluminal flow, which we do not regard as 
a serious flaw, the Sultana-Dyer solution exhibits other  less 
desirable features: the cosmological 
fluid  has negative energy density at late times 
near $R=2M$ \cite{SultanaDyer}; it would be desirable if  the  
cosmological matter  were described by a single fluid composed of 
particles following timelike geodesics instead of an odd mixture
of a null and a massive dust;  and limiting  the scale factor of 
the  universe to the special choice  $a\propto t^{2/3}$ seems too 
restrictive. Work is in progress to find new exact solutions, 
both in general relativity and in alternative theories, which do 
not share these problems.

\section*{Acknowledgments}

Thanks go to the Natural Sciences and  Engineering Research 
Council of Canada  (NSERC) for financial support.

\vskip1truecm
\section*{Appendix~A}
\def\theequation{A.\arabic{equation}}\setcounter{equation}{0}

Eq.~(\ref{ssscurvature}) for the Ricci curvature shows a 
divergence at $\tilde{r}=2m$, which is not noticed in the 
Sultana-Dyer paper. Here we discuss the likely cause of this 
fact. The Ricci scalar coincides (up to the constant $\kappa=8\pi 
G$)  with the negative trace of the 
energy-momentum tensor $T_{ab}$ and the matter energy density and 
(zero) pressure in the Sultana-Dyer paper are identified using 
their Einstein equations~(13) with the stress-energy tensor 
given by their eq.~(4). The latter is meant to provide the 
stress-energy 
tensor $\tilde{T}_{ab}$ obtained by conformally transforming the 
vanishing stress-energy tensor $T_{ab}$ of the Schwarzschild 
solution, which we want to reconsider here. Under the conformal 
transformation of the metric $ g_{ab} \rightarrow 
\tilde{g}_{ab}=\Omega^2 \, g_{ab}$ the Ricci tensor and Ricci 
scalar transform according to
\begin{eqnarray}
&& \tilde{R}_{ab}=R_{ab}-2\nabla_a\nabla_b \ln \Omega -
g_{ab} g^{ef} \nabla_e \nabla_f \ln \Omega \nonumber\\ 
&&\nonumber\\
&& + 2 \left( \nabla_a \ln \Omega \right)
\left( \nabla_b \ln \Omega \right)
-2 g_{ab} g^{ef} \left( \nabla_e \ln \Omega \right)
\left( \nabla_f \ln \Omega \right) \;,\nonumber\\
&&\\
&&\nonumber\\
&& {{\tilde{R}}^a}_a=\Omega^{-2} \left( 
{R^a}_a-\frac{6\Box\Omega}{\Omega} 
\right) \;, \label{pippanera}
\end{eqnarray}
respectively \cite{Wald}. Using the Einstein equations 
$ \tilde{R}_{ab}-\frac{1}{2} \tilde{g}_{ab}\tilde{R}=\kappa \, 
\tilde{T}_{ab}$ and 
$ R_{ab}-\frac{1}{2} g_{ab} R =\kappa \, T_{ab}$, one obtains
\begin{eqnarray} 
&& \kappa\, \tilde{T}_{ab}=\kappa\, 
T_{ab}-\frac{2\nabla_a\nabla_b\Omega}{\Omega^2} 
+\frac{4\nabla_a\Omega\nabla_b\Omega}{\Omega^2} \nonumber\\
&&\nonumber\\
&& -g_{ab} \, \frac{g^{cd}\nabla_c\Omega\nabla_d\Omega}{\Omega^2} 
 +2g_{ab}\,  \frac{g^{cd}\nabla_c\nabla_d\Omega}{\Omega} \;.
\end{eqnarray}
Sultana and Dyer instead have, for vanishing $T_{ab}$ (eq.~(4) 
of \cite{SultanaDyer}),
\be
\kappa\, \tilde{T}_{ab}=2\tilde{g}_{ab} \, \frac{ 
\tilde{\nabla}^2\Omega}{\Omega} - 
\frac{2\tilde{\nabla}_a\tilde{\nabla}_b \Omega}{\Omega} 
-\frac{3}{\Omega^2} \tilde{g}_{ab} 
\tilde{g}^{mn}\tilde{\nabla}_m\Omega \tilde{\nabla}_n\Omega \;.
\ee
The discrepancy between these formulas, plus the fact that the 
tilded 
operator $\tilde{\nabla}$ is not the correct one to be used in 
the 
expression of $\tilde{T}_{ab}$, is likely to be the reason why 
the singularity at $\tilde{r}=2m$ 
is missed in \cite{SultanaDyer}. The correct Ricci scalar can be 
calculated using eq.~(\ref{pippanera}) with ${R^a}_a=0$ and 
$a=\bar{t}^{2/3}=\bar{\eta}^2$ yielding
\begin{eqnarray}
 {{\tilde{R}}^a }_a & = & - \frac{6\Box\Omega}{\Omega^3}=
- \frac{6\Box \left( \bar{\eta}^2 \right)}{\bar{\eta}^6} 
= 2g^{0b} \delta_{0a} \nonumber\\
&&\nonumber\\
&=&\frac{12}{\bar{\eta}^6 \left( 
1-\frac{2m}{\tilde{r}} \right)}=\frac{12}{a^3\left( 
1-\frac{2m}{\tilde{r}} \right)} \;.
\end{eqnarray}
The singularity at $\tilde{r}=2m$ was also noted recently in 
\cite{Sun}.

The Sultana-Dyer solution can be reobtained using as material 
source a single imperfect fluid with a radial (spacelike) energy 
current instead of a two-fluid mixture, as shown in the first of 
Refs.~\cite{McClureDyer}. Eq.~(90) of this work for 
${G^a}_a=-{R^a}_a$ is clearly singular at $\tilde{r}=2m$.

\vskip1truecm

\end{document}